\documentclass{iau} 
\usepackage{graphicx}
\usepackage{amsmath}
\usepackage{float}
\usepackage{subcaption}

\title[Ratio of penumbra to umbra area] 
{Long-term variation of sunspot penumbra to umbra area ratio\\[0.3cm] \small{A study using Kodaikanal white-light Digitized Data}}

\author[Bibhuti \etal]   
{Bibhuti Kumar Jha$^1$, Sudip Mandal $^1$
 \and Dipankar Banerjee$^{1,2}$}

\affiliation{$^1$Indian Institute of Astrophysics, \\ Banglore 560034, India
\\ email: {\tt bibhuti.kj@iiap.res.in} \\[\affilskip]
$^2$Center of Excellence in Space Sciences India, IISER Kolkata,\\ Mohanpur 741246, West Bengal, India
 \\email: {\tt dipu@iiap.res.in}}

\pubyear{2018}
\volume{340}  
\jname{Long-term datasets for the understanding of solar and stellar magnetic cycles}
\editors{D. Banerjee, J. Jiang, K. Kusano \& S. K. Solanki, eds.}
\begin{document}

\maketitle
\begin{abstract}
A typical sunspot, as seen in white-light intensity images, has a two part structure:
 a dark umbra and a lighter penumbra. Such distinction primarily arises due to the different
 orientations of magnetic fields in these two regions. In this study, we use the Kodaikanal 
 white-light digitized data archive to analyze the long-term evolution of umbral and penumbal 
 area. We developed an 'automated  algorithm' to uniquely identify the sunspot umbra (including
 the calculation of penumbra to umbra ratio) from these digitized intensity images. Our analysis 
 reveals that the ratio increases slightly with the increase of sunspot area upto $100~\mu$Hem but 
 eventually settles down to a constant value after that.This study, not only allows us to better understand the evolution 
of an individual spot and  its corresponding magnetic field but this is also beneficial for solar 
dynamo studies which aim to reproduce such structures using a MHD theory.
\end{abstract}

\firstsection 
\section{Introduction}
The earliest study of the Sun revels the umbral-penumbral structure of the sunspots, these are the two region of a sunspot; a darker umbra surrounded by lighter penumbra. Form the recent observations it becomes clear that,
 umbrae are the regions of vertical magnetic field whereas field lines are more inclined in penumbrae.

\cite[Nicholson (1933)]{1933PASP...45...51N} studied the relative size of sunspot penumbra to umbra, derived from Royal Observatory of Greenwich (RGO) data and found that the ratio is fairly constant ($\approx 4.7$). \cite[Waldmeier (1939)]{1939MiZur..14..439W} examined their relative size using the different data set and his result was in agreement with Nicholson (1933). \cite[Antalov{\'a} (1971)]{1971BAICz..22..352A} and
\cite[Hathaway (2013)]{2013SoPh..286..347H} have studied the sunspot penumbral to umbral area ratio using the RGO data and found that the penumbral to umbral area ratio increases with sunspot 
size for smaller sunspots but it is almost independent of the size for large sunspots. \cite[Hathaway (2013)]{2013SoPh..286..347H}   also showed some long term trend in the ratio of penumbra to umbra area for smaller sunspot group.

Study of variation in total solar irradiance uses total sunspot area and assume that the ratio of penumbral to umbral area does not change 
with time (\cite[Foukal \& Lean 1990]{1990Sci...247..556F}). If there is any significant variation in the ratio it will help us to better reconstruct the total solar irradiance. In addition to that, it will also help us to understand penumbral formation in different sizes of sunspot.

 In this article, we discuss an automated algorithm to detect the umbra from the sunspot and examine the variation of penumbral to umbral area ratio using white light intensity images of the Sun obtained from Kodaikanal solar observatory.

\vspace*{-0.5 cm}
\section{Data and Methodology}
Kodaikanal solar observatory has been observing the Sun in three different wavelength bands namely White light (since 1904), Ca-K (since 1905) and  H-alpha (since 1912). These data were taken on photographic plates and recently they have been digitized (4k$\times$4k) and calibrated.  
Sunspots have been  detected from the full disk images using semi automated algorithm named `modified-STARA' (\cite[Ravindra et al. 2013]{2013A&A...550A..19R} \& \cite[Mandal et al. 2017]{2017A&A...601A.106M}) and the binary masks of detected sunspots are stored in Idl save files. We have developed an automatic algorithm in Idl, to detect the umbra  from individual sunspot based on Otsu technique (\cite[Otsu 1979)]{4310076}. The penumbral to umbral area are corrected for foreshortening.  An example is shown in Fig.\,\ref{fig1}(a). Using this algorithm umbra has been detected for all sunspots present in white light images taken from 1923 to 2011 (Cycle 16-23). Penumbral to umbral area ratio for each sunspot has been calculated as 
$$\text{Ratio}=\frac{\text{Total area}}{\text{Area of umbra}}-1$$ 
\section{Results and Conclusion}
\begin{figure}[!h]
\begin{center}
 \includegraphics[scale=0.35]{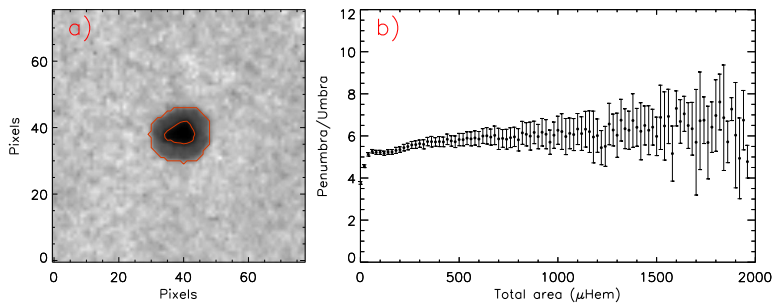} 
 \vspace*{-0.3 cm}
 \caption{a) A representative example of a sunspot and umbra detection. b) Variation of ratio as a function of total sunspot area, averaged over a bin size of 20 $\mu$Hem.}
   \label{fig1}
\end{center}
\end{figure}
\vspace*{-0.3 cm}
The ratio increases with sunspot area for smaller sunspots ($<100~\mu$Hem) whereas it does not show any significant variation with  total area for larger sunspots ($>100~\mu$Hem) (Shown in Fig.\ref{fig1}(b)). This suggest that, as the  size of sunspot increases, there must be the emergence of additional penumbra in the spot. Average ratio for all sunspots is $\approx 5.5$ and this result is in agreement with earlier results
 (\cite[Antalov{\'a} 1971]{1971BAICz..22..352A} \& \cite[Hathaway 2013]{2013SoPh..286..347H}).


\vspace{-0.4cm}
\begin{thebibliography}{}


\bibitem[Antalov{\'a} (1971)]{1971BAICz..22..352A}
{Antalov{\'a}}, A. 1971, 
\textit{Bulletin of the Astronomical Institutes of Czechoslovakia}, 22, 352

\bibitem[Foukal \& Lean (1990)]{1990Sci...247..556F}
{Foukal}, P., \& {Lean},J. 1990, 
\textit{Science}, 247, 556


\bibitem[Hathaway (2013)]{2013SoPh..286..347H}
{Hathaway}, D.~H. 2013,
\textit{Solar Phys.}, 286, 347

\bibitem[Mandal \etal (2017)]{2017A&A...601A.106M}
{Mandal}, S., {Hegde}, M., {Samanta}, T., {Hazra}, G., {Banerjee}, D., \&
  {Ravindra}, B. 2017,
   \textit{A\&A}, 601, A106

\bibitem[Nicholson (1933)]{1933PASP...45...51N}
{Nicholson}, S.~B. 1933,
 \textit{Publ. Astron. Soc. Pac.}, 45, 51

\bibitem[Otsu 1979)]{4310076}
Otsu, N. 1979,
 \textit{IEEE Transactions on Systems, Man, and Cybernetics}, 9, 62

\bibitem[Ravindra \etal (2013)]{2013A&A...550A..19R}
{Ravindra}, B., {Priya}, T.~G., {Amareswari}, K., {Priyal}, M., {Nazia}, A.~A.,
  \& {Banerjee}, D. 2013,
  \textit{A\&A}, 550, A19

\bibitem[Waldmeier (1939)]{1939MiZur..14..439W}
{Waldmeier}, M. 1939,
 \textit{Astronomische Mitteilungen der Eidgen{\"o}ssischen
  Sternwarte Zurich}, 14, 439
  













\end{thebibliography}
\end{document}